\documentclass[aps,prl,preprint,groupedaddress]{revtex4-1}
\usepackage{amssymb}
\usepackage{amsmath}
\usepackage{bbold}
\usepackage{slashed}
\usepackage{IEEEtrantools}
\usepackage{physics}
\usepackage{bm}
\usepackage{longtable}
\usepackage{graphicx}
\usepackage{subfigure}

\begin{document}


\title{Estimation on the neutrino masses by using neutrino oscillation parameters}


\author{Wen-Jie Wu}
\author{Xiang Zhou}
\email[]{xiangzhou@whu.edu.cn}
\affiliation{Hubei Nuclear Solid Physics Key Laboratory, School of Physics and Technology, Wuhan University, Wuhan 430072, China}


\date{\today}

\begin{abstract}
	It has been shown that neutrino masses can be determined under the particle ansatz. In this paper, we give the general formulas of neutrino masses related to the neutrino oscillation parameters which show that there is a mass hierarchy transition. Using the best fit values measured by electron and muon neutrino oscillation experiments, the total neutrino mass is about 0.25 eV. According to the standard neutrino cosmology, the neutrino energy density is about 55 eV/cm$^3$ and $\Omega_m h^2>\Omega_\nu h^2\sim 0.0055$. 
\end{abstract}


\maketitle

Nonzero neutrino masses are beyond the standard model which are confirmed by various oscillation experiments. The phase difference (PhD) in the oscillation probability derived under the so called ``same energy'' or ``same momentum'' ansatz in the traditional phenomenological theory of neutrino oscillations\cite{Thomson} is
\begin{equation}
	\Delta\phi_{ji}\simeq\frac{m_j^2-m_i^2}{4E_\nu}L=\frac{\Delta m_{ji}^2}{4E_\nu}L,
	\label{oldPhD}
\end{equation}
where $m_i$ and $m_j$ are the mass eigenvalues, $E_\nu$ is the neutrino energy, $L$ is the neutrino flight distance. It is only sensitive to the mass square difference $\Delta m_{ji}^2$. However, the so called ``particle'' ansatz assumes that in the rest frame of a flavor neutrino $\nu_\alpha$ all momentums of its mass eigenstates $\ket{\nu_i}$ are simultaneously zero\cite{ZhouXmmPhD}. Under the particle ansatz, the phase difference $\Delta\phi_{ji}$ is mass measurable which is
\begin{equation}
	\Delta\phi_{ji} \simeq \frac{2m_{\nu_\alpha}(m_j-m_i)}{4 E_{\nu_\alpha}}L=\frac{\Delta m^2_{\alpha ji}}{4 E_{\nu_\alpha}}L,
                    \label{mmPhD}
\end{equation}
where $L$ is the distance and $\Delta m^2_{\alpha ji}=2m_{\nu_\alpha}(m_j-m_i)$. In this paper, we give the general formulas of the neutrino masses related to the neutrino oscillation parameters, and estimate the total mass and the neutrino energy density of the Universe. 

In the case of 3-neutrino mixing $U_{\alpha i}$ is the element of the PMNS matrix $U=[U_{\alpha i}]$ with three mixing angles $\theta_{12}$, $\theta_{23}$ and $\theta_{13}$ and Dirac phase $\delta$. The PMNS matrix can be written as
\begin{equation}
	U_\mathrm{PMNS}=
	\begin{pmatrix}
		U_{e 1}    & U_{e 2}    & U_{e 3}    \\
		U_{\mu 1}  & U_{\mu 2}  & U_{\mu 3}  \\
		U_{\tau 1} & U_{\tau 2} & U_{\tau 3} \\
	\end{pmatrix}=
	\begin{pmatrix}
		1 & 0       & 0      \\
		0 &  c_{23} & s_{23} \\
		0 & -s_{23} & c_{23} \\
	\end{pmatrix}
	\begin{pmatrix}
		c_{13}            & 0 & s_{13}e^{-i\delta} \\
		0                 & 1 & 0                  \\
	       -s_{13}e^{i\delta} & 0 & c_{13}             \\
	\end{pmatrix}
	\begin{pmatrix}
		c_{12} & s_{12} & 0 \\
	       -s_{12} & c_{12} & 0 \\
		0      & 0      & 1 \\
	\end{pmatrix},
\end{equation}
where $c_{ij}=\cos\theta_{ij}$ and $s_{ij}=\sin\theta_{ij}$. $m_{\nu_\alpha}$ and $m_i$ have the relationship as
\begin{eqnarray}
\label{eq:NortransformMatrix}
       \begin{pmatrix}
               m_{\nu_e}    \\
               m_{\nu_\mu}  \\
               m_{\nu_\tau} 
       \end{pmatrix}=
       \begin{pmatrix}
               |U_{e 1}|^2    & |U_{e 2}|^2    & |U_{e 3}|^2    \\
               |U_{\mu 1}|^2  & |U_{\mu 2}|^2  & |U_{\mu 3}|^2  \\
               |U_{\tau 1}|^2 & |U_{\tau 2}|^2 & |U_{\tau 3}|^2 
       \end{pmatrix}
       \begin{pmatrix}
               m_1 \\
               m_2 \\
               m_3
       \end{pmatrix}.
\end{eqnarray}
\begin{equation}
\label{eq:InvtransformMatrix}
       \begin{pmatrix}
               m_1 \\
               m_2 \\
               m_3
       \end{pmatrix}=
       \begin{pmatrix}
               \overline{U^2}_{e 1} & \overline{U^2}_{\mu 1} & \overline{U^2}_{\tau 1}    \\
               \overline{U^2}_{e 2} & \overline{U^2}_{\mu 2} & \overline{U^2}_{\tau 2}    \\
               \overline{U^2}_{e 3} & \overline{U^2}_{\mu 3} & \overline{U^2}_{\tau 3}
        \end{pmatrix}
        \begin{pmatrix}
                m_{\nu_e}    \\
                m_{\nu_\mu}  \\
                m_{\nu_\tau}
        \end{pmatrix},
\end{equation}
where 
\begin{equation}
	\begin{pmatrix}
               |U_{e 1}|^2    & |U_{e 2}|^2    & |U_{e 3}|^2    \\
               |U_{\mu 1}|^2  & |U_{\mu 2}|^2  & |U_{\mu 3}|^2  \\
               |U_{\tau 1}|^2 & |U_{\tau 2}|^2 & |U_{\tau 3}|^2
       \end{pmatrix}
	\begin{pmatrix}
               \overline{U^2}_{e 1} & \overline{U^2}_{\mu 1} & \overline{U^2}_{\tau 1}    \\
               \overline{U^2}_{e 2} & \overline{U^2}_{\mu 2} & \overline{U^2}_{\tau 2}    \\
               \overline{U^2}_{e 3} & \overline{U^2}_{\mu 3} & \overline{U^2}_{\tau 3}
        \end{pmatrix}=
	\begin{pmatrix}
		1 & 0 & 0 \\
		0 & 1 & 0 \\
		0 & 0 & 1
	\end{pmatrix}.
\end{equation}
There are five independent conditions for $\overline{U^2}_{\alpha i}$ which are 
\begin{eqnarray}
	\overline{U^2}_{e 1}+\overline{U^2}_{\mu 1}+\overline{U^2}_{\tau 1}  & = & 1, \nonumber\\
	\overline{U^2}_{e 2}+\overline{U^2}_{\mu 2}+\overline{U^2}_{\tau 2}  & = & 1, \nonumber\\
	\overline{U^2}_{e 3}+\overline{U^2}_{\mu 3}+\overline{U^2}_{\tau 3}  & = & 1, \nonumber\\
	\overline{U^2}_{e 1}+\overline{U^2}_{e 2}+\overline{U^2}_{e 3}       & = & 1, \nonumber\\
	\overline{U^2}_{\mu 1}+\overline{U^2}_{\mu 2}+\overline{U^2}_{\mu 3} & = & 1. 
\end{eqnarray}
Therefore, there are only four independent elements in the matrix $\overline{U^2}=[\overline{U^2}_{\alpha i}]$. In this paper, $\overline{U^2}_{e 1}$, $\overline{U^2}_{\mu 1}$, $\overline{U^2}_{e 2}$ and $\overline{U^2}_{\mu 2}$ are chosen. 

The total neutrino mass $M$ satisfies 
\begin{equation}
	M=m_{\nu_e}+m_{\nu_\mu}+m_{\nu_\tau}=m_1+m_2+m_3.
\end{equation}
If the coefficients $k=m_{\nu_e}/M$ and $\epsilon=m_{\nu_\mu}/m_{\nu_e}$ are introduced, the masses of flavor neutrinos can be written as
\begin{eqnarray}
	m_{\nu_e} & = & kM, \nonumber\\
	m_{\nu_\mu} & = & \epsilon kM, \nonumber\\
	m_{\nu_\tau} & = & (1-k-\epsilon k)M, \nonumber
\end{eqnarray}
where $0<k<1$, $0<\epsilon k<1$ and $0<1-k-\epsilon k<1$. For neutrino experiments a traditional mass square term in the PhD under the ``same energy'' or ``same momentum'' ansatz can be transformed to its corresponding mmPhD which are listed in Table~\ref{tab:deltamsqure}. The value of $\epsilon=m_{\nu_\mu}/m_{\nu_e}=|\Delta m^2_{\mu 32}|/|\Delta m^2_{e32}|$ can be obtained by spectral analyses under the particle ansatz. 
\begin{table}
\caption{\label{tab:deltamsqure} The correspondence of mass square differences between the ``same energy" ansatz (or ``same momentum" ansatz) and the "particle ansatz".}
\begin{ruledtabular}
\begin{tabular}{ccclll}
	Probability                        & Mass square term in the    & Mass square term in \\
	                                   & ``same energy'' ansatz or & the particle ansatz \\
				           & ``same momentum'' ansatz  & \\
\colrule
	$P(\nu_e\to\nu_e)$                 & $\Delta m_{21}^2$         & $\Delta m^2_{e21} = 2m_{\nu_e}(m_2-m_1)                 $\\
	$P(\bar{\nu}_e\to\bar{\nu}_e)$     &                           & $\phantom{\Delta m^2_{e21}} = 2m_{\bar{\nu}_e}(m_2-m_1) $\\
\colrule
	$P(\nu_\mu\to\nu_\mu)$             & $|\Delta m_{32}^2|$       & $|\Delta m^2_{\mu 32}|=2m_{\nu_\mu}|m_3-m_2|$      \\
	$P(\bar{\nu}_\mu\to\bar{\nu}_\mu)$ &                           & $\phantom{\Delta m^2_{\mu 32}} = 2m_{\bar{\nu}_\mu}|m_3-m_2|$\\
\colrule
	$P(\nu_e\to\nu_e)$                 & $|\Delta m_{32}^2|$       & $|\Delta m^2_{e32}| = 2m_{\nu_e}|m_3-m_2|$        \\
	$P(\bar{\nu}_e\to\bar{\nu}_e)$     &                           & $\phantom{\Delta m^2_{e32}} = 2m_{\bar{\nu}_e}|m_3-m_2|$  \\
\end{tabular}
\end{ruledtabular}
\end{table}

$\Delta m^2_{e21}$ and $\Delta m^2_{\mu 32}$ can be rewritten as

\begin{eqnarray}
    \label{me21}
	\Delta m^2_{e21} & = & 2[(\overline{U^2}_{e2}-\overline{U^2}_{e1})k^2                              \nonumber\\
			 &   & +(\overline{U^2}_{\mu 2}-\overline{U^2}_{\mu 1})\epsilon k^2                         \\
			 &   & +(\overline{U^2}_{e1}+\overline{U^2}_{\mu 1}-\overline{U^2}_{e2}-\overline{U^2}_{\mu 2})(k-k^2-\epsilon k^2)]M^2, \nonumber                                  
\end{eqnarray}
\begin{eqnarray}
    \label{mmu32}
	|\Delta m^2_{\mu 32}| & = & 2|(1-\overline{U^2}_{e1}-2\overline{U^2}_{e2})\epsilon k^2       \nonumber\\
			    &   & +(1-\overline{U^2}_{\mu 1}-2\overline{U^2}_{\mu 2})\epsilon^2k^2          \\
			    &   & +(\overline{U^2}_{e1}+2\overline{U^2}_{e2}+\overline{U^2}_{\mu 1}+2\overline{U^2}_{\mu 2}-2)(\epsilon k-\epsilon k^2-\epsilon^2 k^2)|M^2. \nonumber
\end{eqnarray}
It is easy to show that if $\epsilon=1$, $k=1/3$ and then $M\to\infty$. Thus the masses of three flavor neutrinos $\nu_{\alpha}$ can not be all the same. $\eta$ is defined as the ratio of $\Delta m^2_{e 21}$ and $\Delta m^2_{\mu 32}$ which is
\begin{equation}
\label{eq:eta}
\eta\equiv\frac{\Delta m^2_{e21}}{|\Delta m^2_{\mu 32}|}=\frac{A}{|B|},
\end{equation}
where
\begin{eqnarray}
	A & \equiv & \overline{U^2}_{e1}+\overline{U^2}_{\mu 1}-\overline{U^2}_{e2}-\overline{U^2}_{\mu 2} \nonumber\\
	  &        & +(2\overline{U^2}_{e2}-2\overline{U^2}_{e1}+\overline{U^2}_{\mu 2}-\overline{U^2}_{\mu 1} \nonumber\\
	  &        &+(2\overline{U^2}_{\mu 2}-2\overline{U^2}_{\mu 1}+\overline{U^2}_{e2}-\overline{U^2}_{e1})\epsilon)k \nonumber\\
	B & \equiv & (\overline{U^2}_{e1}+2\overline{U^2}_{e2}+\overline{U^2}_{\mu 1}+2\overline{U^2}_{\mu 2}-2) \epsilon\nonumber\\
	  &        & +((3-2\overline{U^2}_{e1}-4\overline{U^2}_{e2}-\overline{U^2}_{\mu 1}-2\overline{U^2}_{\mu 2})\epsilon \nonumber\\
	  &        & +(3-2\overline{U^2}_{\mu 1}-4\overline{U^2}_{\mu 2}-\overline{U^2}_{e1}-2\overline{U^2}_{e2})\epsilon^2)k \nonumber
\end{eqnarray}

\begin{figure}
\caption{Mass hierarchy transition where the red region is for normal hierarchy and the blue one is for inverted hierarchy.}
\label{fig:MHT}
\includegraphics[width=0.5\textwidth]{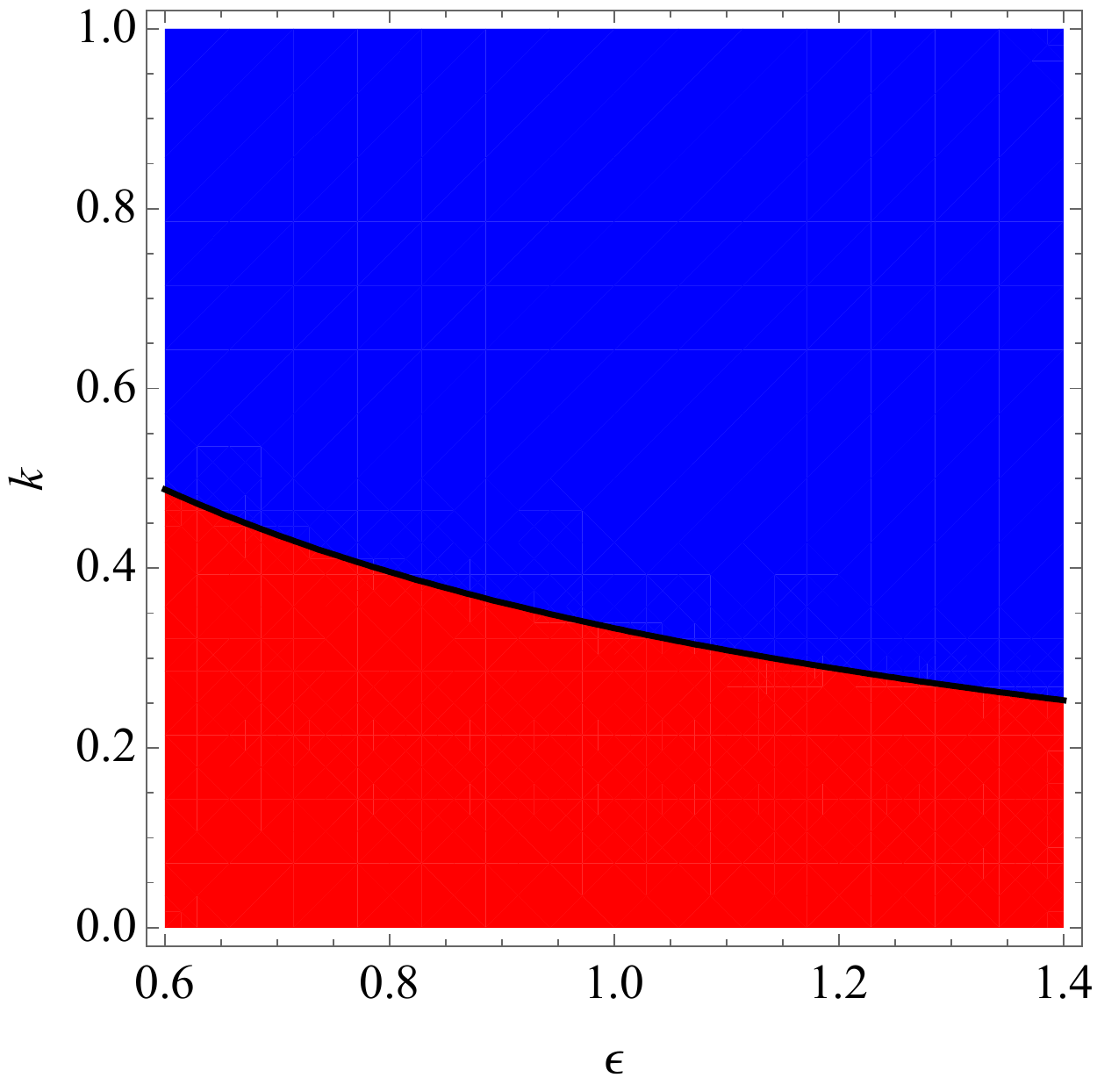}
\end{figure}

Combining the results of mixing angles and mass square differences, $k$ and $M$ can be expressed as Eq.~(\ref{eq:kCoeff}) and Eq.~(\ref{eq:totMass}),
\begin{eqnarray}
	k & = & -{(\overline{U^2}_{e1}+\overline{U^2}_{\mu 1}-\overline{U^2}_{e2}-\overline{U^2}_{\mu 2}\mp(\overline{U^2}_{e1}+2\overline{U^2}_{e2}+\overline{U^2}_{\mu 1}+2\overline{U^2}_{\mu 2}-2)\eta\epsilon)} \nonumber\\
	  &   & /(2\overline{U^2}_{e2}-2\overline{U^2}_{e1}+\overline{U^2}_{\mu 2}-\overline{U^2}_{\mu 1}+(2(1\pm\eta)\overline{U^2}_{\mu 2}-(2\mp\eta)\overline{U^2}_{\mu 1}+(1\pm 4\eta)\overline{U^2}_{e2} \nonumber\\
	  &   & -(1\mp 2\eta)\overline{U^2}_{e1}\pm 3\eta)\epsilon\mp(3-2\overline{U^2}_{\mu 1}-4\overline{U^2}_{\mu 2}-\overline{U^2}_{e1}-2\overline{U^2}_{e2})\eta\epsilon^2),
	\label{eq:kCoeff} \\
	M & = & \{2[(\overline{U^2}_{e2}-\overline{U^2}_{e1})k^2+(\overline{U^2}_{\mu 2}-\overline{U^2}_{\mu 1})\epsilon k^2 \nonumber \\
	  &   & +(\overline{U^2}_{e1}+\overline{U^2}_{\mu 1}-\overline{U^2}_{e2}-\overline{U^2}_{\mu 2})(k-k^2-\epsilon k^2)]/\Delta m^2_{e21}\}^{-\frac{1}{2}},
	\label{eq:totMass}
\end{eqnarray}
where the upper operator in $\mp$ or $\pm$ is for $B>0$ and the lower operator for $B<0$. If the parameters of the PMNS matrix are known and $\epsilon=m_{\nu_\mu}/m_{\nu_e}$ is obtained by measuring the ratio of $m_{\mu 32}$ and $m_{e 32}$ from neutrino experiments, then the fraction factor $k=m_{\nu_e}/M$ and the total flavor neutrinos mass $M$ can be calculated from Eq.~(\ref{eq:kCoeff}) and Eq.~(\ref{eq:totMass}). Consequently all  neutrino masses of flavor and mass eigenstate can be obtained. Fig.~\ref{fig:MHT} shows that there is a mass hierarchy transition with $\theta_{12}=34.7^\circ$, $\theta_{13}=8.43^\circ$, $\theta_{23}=30^\circ$ and $\delta=\pm 90^\circ$.    

\begin{figure}
\caption{The effects of Dirac phase $\delta$ and mixing angle $\theta_{23}$ on the coefficient $k$ and the total neutrino mass $M$. The mass hierarchy of both parameter set (I) and (II) are normal in the whole ($\delta$, $\theta_{23}$) parameter space.}
\label{fig:kM}
\subfigure[~Parameter set (I), $k(\delta,\theta_{23})$] {
    {\label{fig:Ik}\includegraphics[width=0.48\textwidth]{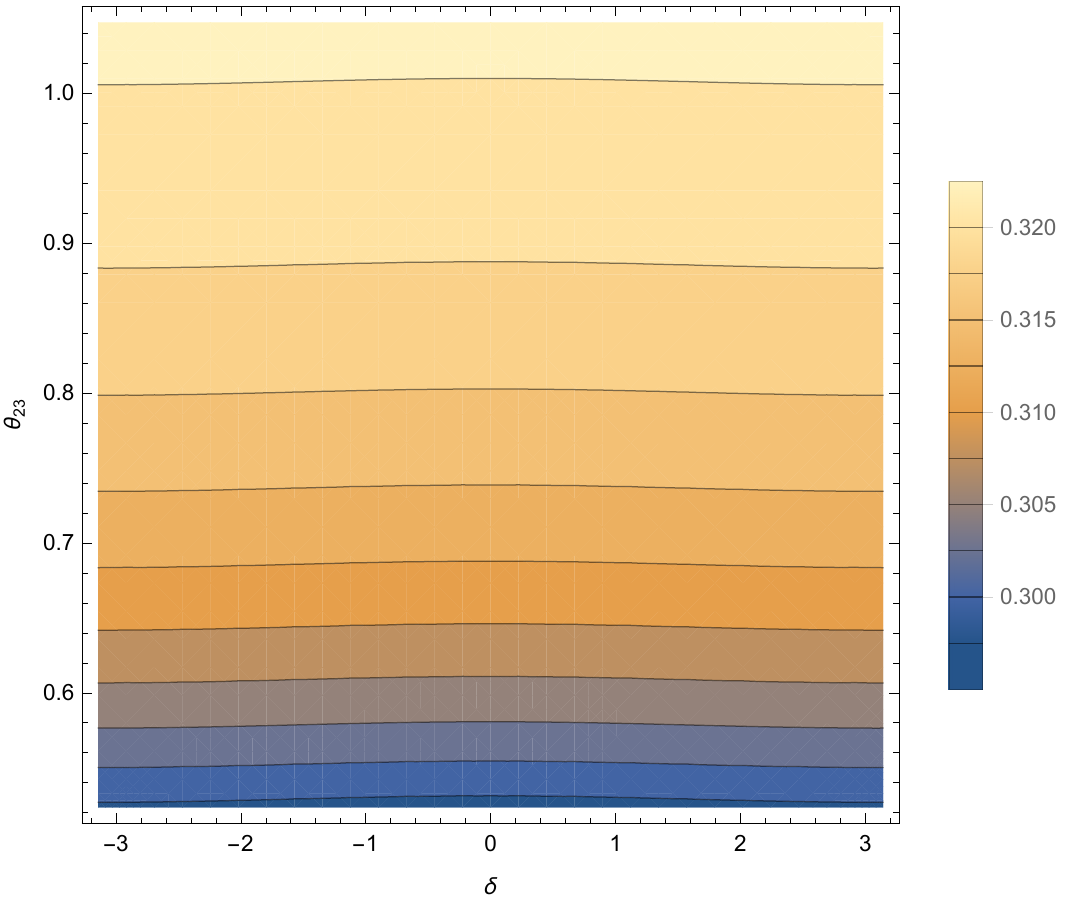}}}
\subfigure[~Parameter set (I), $M(\delta,\theta_{23})$] {
    {\label{fig:IM}\includegraphics[width=0.48\textwidth]{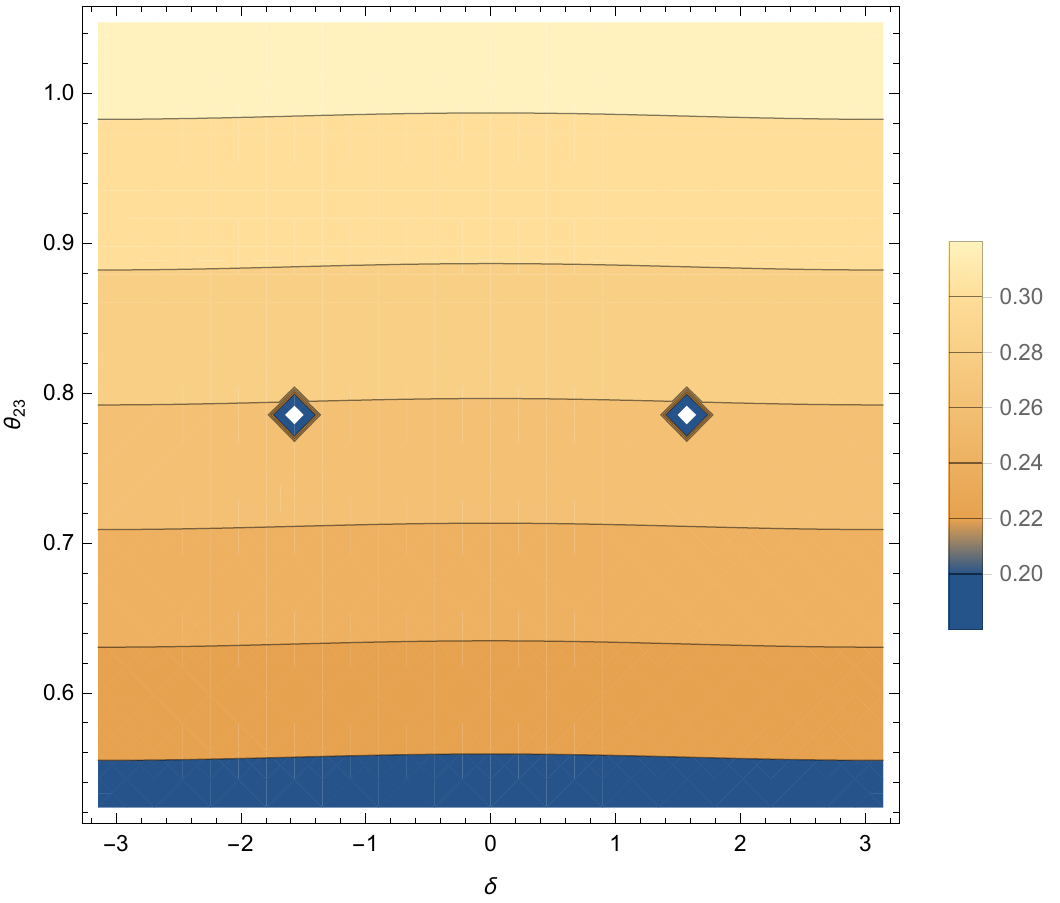}}}
\subfigure[~Parameter set (II), $k(\delta,\theta_{23})$] {
    {\label{fig:IIk}\includegraphics[width=0.48\textwidth]{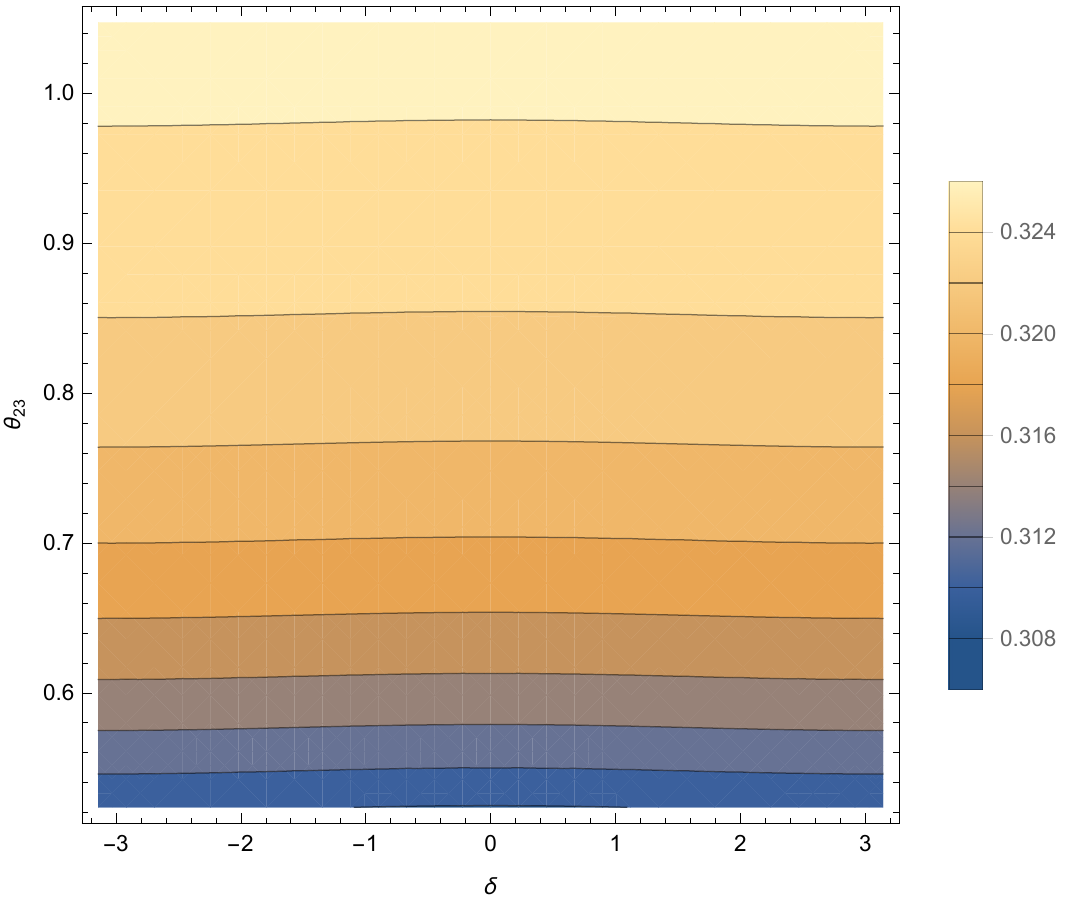}}}
\subfigure[~Parameter set (II), $M(\delta,\theta_{23})$] {
    {\label{fig:IIM}\includegraphics[width=0.48\textwidth]{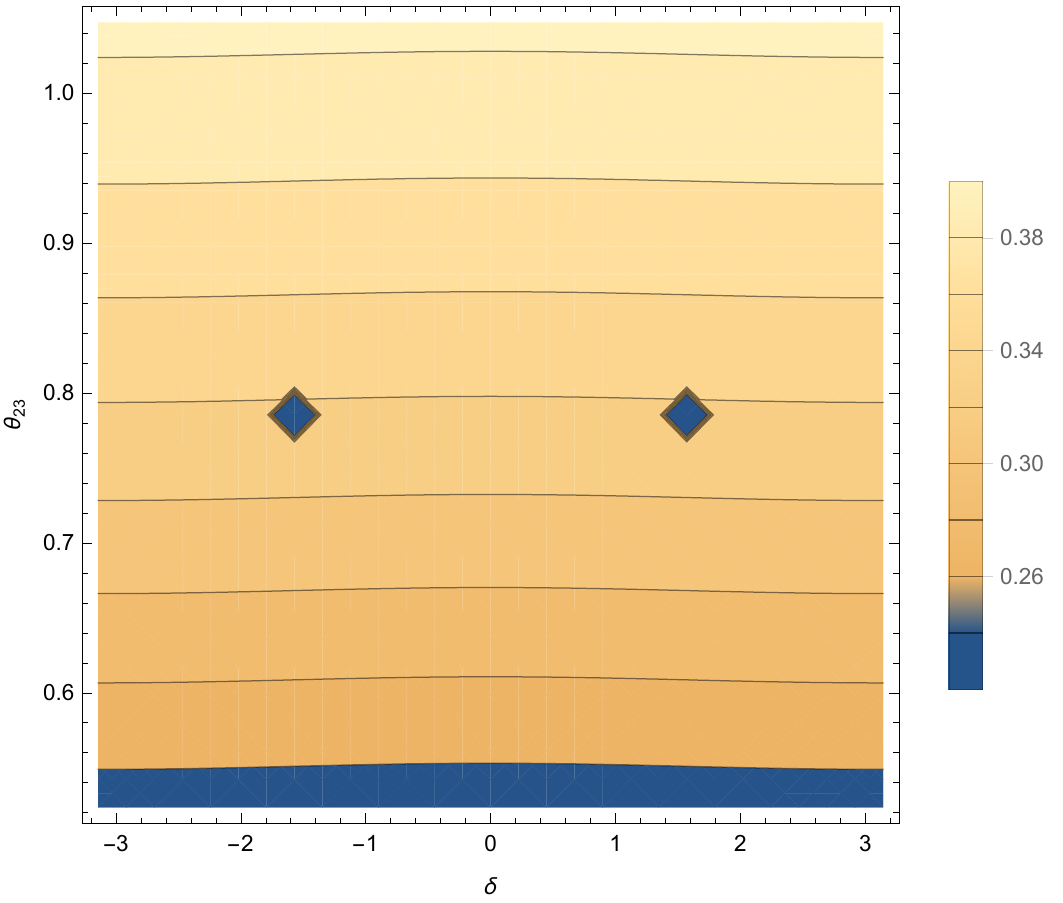}}}
\end{figure}

The KamLAND experiment gave values of $\Delta m_{21}^2$ and $\theta_{12}$ by measuring long baseline reactor neutrinos. The fit values with only KamLAND data are $\Delta m_{21}^2=(7.54^{+0.19}_{-0.18})\times 10^{-5}~\mathrm{eV}^2$ and $\mathrm{tan}^2 \theta_{12}=0.481^{+0.092}_{-0.080}$\cite{KamLAND2014}. The latest Daya Bay results based on 1230 live days of data gave $\mathrm{sin}^2 2\theta_{13}=0.0841\pm 0.0027\pm 0.0019$ and $\Delta m_{32}^2=(2.45\pm 0.06\pm 0.06)\times 10^{-3}~\mathrm{eV}^2$ for normal hierarchy (NH) and $\Delta m_{32}^2=-(2.56\pm 0.06\pm 0.06)\times 10^{-3}~\mathrm{eV}^2$ for inverted hierarchy (IH)\cite{DayaBay2017}. Recently, NO$\nu$A made a precise measurement of $\nu_{\mu}$ disappearance and gave $\Delta m_{32}^2=(+2.67\pm 0.11)\times 10^{-3}~\mathrm{eV}^2$ for NH, and $\Delta m_{32}^2=(-2.72\pm 0.11)\times 10^{-3}~\mathrm{eV}^2$ for IH\cite{NOVA2017}. Based on the parameters derived under the ``same energy" or ``same momentum" ansatz, the center values of corresponding parameters under particle ansatz are listed in Table~\ref{tab:params}. There are two sets of parameters due to the mass hierarchy under the ``same energy" or ``same momentum" ansatz which are distinguished using label (I) and (II) under particle ansatz. In addition, $\epsilon$ and $\eta$ can be obtained and results are listed in table~\ref{tab:params}. With $\theta_{12}$, $\theta_{13}$, $\epsilon$ and $\eta$ in Table~\ref{tab:params}, Fig.~{\ref{fig:kM}} gives $k$ and $M$ of the parameter set (I) and (II)when $\delta$ runs from $-180^\circ$ to $180^\circ$ and $\theta_{23}$ runs from $30^\circ$ to $60^\circ$ according to Eq.~(\ref{eq:kCoeff}) and Eq.~(\ref{eq:totMass}). 
\begin{table}
\caption{\label{tab:params} Center values of oscillation parameters used in the calculation.}
\begin{ruledtabular}
\begin{tabular}{ccclll} 
	Parameters under ``same energy" ansatz     & Source    & Parameters under particle ansatz \\
	or ``same momentum" ansatz & & \\
\colrule
	$\theta_{12}=34.7^{\circ}$                     & KamLAND\cite{KamLAND2014}  & $\theta_{12}=34.7^{\circ}$  \\ 
\colrule
	$\theta_{13}=8.43^{\circ}$                     & Daya Bay\cite{DayaBay2017} & $\theta_{13}=8.43^{\circ}$\\
\colrule
	$\Delta m_{21}^2=7.54\times 10^{-5}~\mathrm{eV}^2$    & KamLAND\cite{KamLAND2014} & $\Delta m_{e21}^2=7.54\times 10^{-5}~\mathrm{eV}^2$  \\ 
\colrule
	$\Delta m_{32}^2=2.45\times 10^{-3}~\mathrm{eV}^2$ (NH)    & Daya Bay\cite{DayaBay2017} & $\Delta m_{e32}^2=2.45\times 10^{-3}~\mathrm{eV}^2$ (I)  \\
	$\Delta m_{32}^2=-2.56\times 10^{-3}~\mathrm{eV}^2$ (IH)   &                            & $\Delta m_{e32}^2=-2.56\times 10^{-3}~\mathrm{eV}^2$ (II) \\      
\colrule
	$\Delta m_{32}^2=2.67\times 10^{-3}~\mathrm{eV}^2$ (NH)    & NO$\nu$A\cite{NOVA2017}    & $\Delta m_{\mu 32}^2=2.67\times 10^{-3}~\mathrm{eV}^2$ (I) \\
	$\Delta m_{32}^2=-2.72\times 10^{-3}~\mathrm{eV}^2$ (IH)   &                            & $\Delta m_{\mu 32}^2=-2.72\times 10^{-3}~\mathrm{eV}^2$ (II)\\   
\colrule
	                               & This work & $\epsilon=1.09$ (I) \\
	                               & $\Delta m^2_{\mu 32}/\Delta m^2_{e32}$ & $\epsilon=1.06$ (II) \\   
\colrule
	                              & This work  & $\eta=0.0282$ (I) \\
	                              & $\Delta m^2_{e21}/\Delta m^2_{\mu 32}$& $\eta=0.0277$ (II) \\   
\end{tabular}
\end{ruledtabular}
\end{table}

We define the compatibility of mass hierarchy so that the mass hierarchy is compatible when the mass hierarchy under the different ansatzs are the same, and incompatible when the mass hierarchy under the different ansatzs are different. It can be shown that the mass hierarchy of the parameter set (I) is compatible but the mass hierarchy of the parameter set (II) is incompatible. The estimation of the neutrino masses are only done for the mass hierarchy compatible parameter set (I). NO$\nu$A gave two statistically degenerate values of $\sin^2 \theta_{23}$ and $\delta$ both in the NH: $\sin^2 \theta_{23}=0.404$, $\delta=1.48\pi$ and $\sin^2 \theta_{23}=0.623$, $\delta=0.74\pi$\cite{NOVA2017PRL2}. The latest T2K results show that $\sin^2 \theta_{23}=0.532$ and $\delta=-1.791$ for NH\cite{T2K2017}. Combining the knowledge of $\theta_{23}$ and $\delta$ with parameter set (I), the absolute neutrino masses can be estimated and shown in Table~\ref{tab:masses}. According to the standard neutrino cosmology, the ratio of neutrinos to photons is 3/11. Therefore, the contribution to the total energy density of the Universe from neutrinos is given by $\rho_{\nu}=m_{\mathrm{tot}}n_{\nu}=m_{\mathrm{tot}}(3/11)n_{\gamma}$, where $n_{\gamma}=412~\mathrm{cm}^{-3}$, $m_{\mathrm{tot}}=\sum_{\nu}(g_{\nu}/2)m_{\nu}$, $g_{\nu}=4$ for neutrinos with Dirac masses and $g_{\nu}=2$ for Majorana neutrinos. The physical density of neutrinos in units of the critical density $\Omega_\nu=\rho_\nu/\rho_c$ can be written as $\Omega_\nu h^2=m_{\mathrm{tot}}/(94~\mathrm{eV})$, where $h$ is the Hubble constant in units of 100 km$\cdot$s$^{-1}$$\cdot$Mpc$^{-1}$\cite{PDG2016}. The neutrino energy density of the Universe with the Dirac neutrino assumption are listed in Table~\ref{tab:masses}. The results show that $\Omega_m h^2>\Omega_\nu h^2\sim0.0055$.
\begin{table}
\caption{\label{tab:masses}The estimation on the neutrino masses by using neutrino oscillation parameters.}
\begin{ruledtabular}
\begin{tabular}{ccc|cccccccc}
  P.S. & $\theta_{23}$ (NO$\nu$A) & $\delta$ (NO$\nu$A) & MH & k & M (eV) & $m_{1}$ (eV) & $m_{2}$ (eV) & $m_{3}$ (eV) & $\rho_{\nu}$ (eV/cm$^3$) & $\Omega_{\nu}h^2$ \\
\colrule
  (I) & 39.5$^\circ$ & 1.48$\pi$ rad & NH & 0.310 & 0.235 & 0.0722 & 0.0728 & 0.0896 & 52.8 & 0.00499 \\
  (I) & 52.1$^\circ$ & 0.74$\pi$ rad & NH & 0.318 & 0.286 & 0.0904 & 0.0908 & 0.104 & 64.3 & 0.00609 \\
\colrule 
  P.S. & $\theta_{23}$ (T2K) & $\delta$ (T2K) & MH & k & M (eV) & $m_{1}$ (eV) & $m_{2}$ (eV) & $m_{3}$ (eV) & $\rho_{\nu}$ (eV/cm$^3$) & $\Omega_{\nu}h^2$ \\
\colrule
  (I) & 46.8$^\circ$ & -1.791 rad & NH & 0.316 & 0.265 & 0.0833 & 0.0837 & 0.0984 & 59.6 & 0.00564 \\          
\end{tabular}
\end{ruledtabular}
\end{table}

We have estimated the neutrino masses based on the neutrino oscillation parameters\cite{DayaBay2017,NOVA2017,NOVA2017PRL2,T2K2017,PDG2016}. The total neutrino mass is about 0.25 eV and the neutrino energy density of the Universe is about 55 eV/cm$^3$ and then $\Omega_m h^2>\Omega_\nu h^2\sim0.0055$. In the future the precise values with their uncertainties will be determined under the particle ansatz by the spectral analyses of neutrino oscillation experiments\cite{PDG2016,JUNO2016}.    

\appendix*
\section{Appendix A: Formulas of $\overline{U^2}_{e 1}$, $\overline{U^2}_{\mu 1}$, $\overline{U^2}_{e 2}$ and $\overline{U^2}_{\mu 2}$.}
\begin{eqnarray}
    \overline{U^2}_{e 1} & = & \frac{4 \cos\theta _{12} \cos ^2\theta _{13}\cos \theta _{12} \cos 2 \theta _{23}-\cos \delta  \sin \theta _{12}\sin \theta _{13}\sin 2 \theta _{23}}{\cos \delta  \sin 2 \theta _{12} \sin \theta _{13} \sin 2 \theta _{23} -3 \cos 2 \theta _{13}+4 \cos 2 \theta _{12} \cos 2 \theta _{13} \cos 2 \theta _{23}} \nonumber \\
    \overline{U^2}_{\mu 1} & = & (2 \sin ^2\theta _{13} \cos \delta  \sin 2 \theta _{12} \sin 2 \theta _{23} \sin \theta _{13}+2 \sin ^2\theta _{12} \sin ^2\theta _{13} \cos ^2\theta _{23} \nonumber \\
    &&+2 \sin ^2\theta _{23} \cos ^2\theta _{12}-4 \sin ^2\theta _{12} \cos ^4\theta _{13} \cos ^2\theta _{23}) \nonumber \\
    &&/(\cos \delta  \sin 2 \theta _{12} \sin \theta _{13} \sin 2 \theta _{23} -3 \cos 2 \theta _{13}+4 \cos 2 \theta _{12} \cos 2 \theta _{13} \cos 2 \theta _{23}) \\
    \overline{U^2}_{e 2} & = &-\frac{4 \sin \theta _{12} \cos ^2\theta _{13} \cos \delta  \sin \theta _{13} \sin 2 \theta _{23} \cos \theta _{12}+\sin \theta _{12} \cos 2 \theta _{23}}{\cos \delta  \sin 2 \theta _{12} \sin \theta _{13} \sin 2 \theta _{23} -3 \cos 2 \theta _{13}+4 \cos 2 \theta _{12} \cos 2 \theta _{13} \cos 2 \theta _{23}}  \nonumber \\
    \overline{U^2}_{\mu 2} & = &\frac{2 \sin ^2\theta _{13} \cos \delta  \sin 2 \theta _{12} \sin \theta _{13} \sin 2 \theta _{23}-2 \sin ^2\theta _{12} \sin ^2\theta _{23}+4 \cos ^2\theta _{12} \cos 2 \theta _{13} \cos ^2\theta _{23}}{\cos \delta  \sin 2 \theta _{12} \sin \theta _{13} \sin 2 \theta _{23} -3 \cos 2 \theta _{13}+4 \cos 2 \theta _{12} \cos 2 \theta _{13} \cos 2 \theta _{23}} \nonumber
\end{eqnarray}

\begin{acknowledgments}
	X. Z. thanks the helpful discussions of Dr. Zhenyu Zhang and Dr. Qian Liu. This work is supported by the Major Program of the National Natural Science Foundation of China (Grant No. 11390381).
\end{acknowledgments}


\end{document}